\documentclass{aa}  

\usepackage{graphicx}
\usepackage[varg]{newtxmath}
\usepackage{newtxtext}
\usepackage{booktabs}
\usepackage{xcolor}
\def\pd#1#2{\frac{\partial #1}{\partial #2}}

\begin{document}

   \title{The impact of thermal winds on the outburst lightcurves of black hole X-ray binaries}
   \author{Guillaume Dubus
		\inst{1}
		\and
		Chris Done
		\inst{2}
		\and
		Bailey E. Tetarenko
		\inst{3}
		\and
		Jean-Marie Hameury
		\inst{4}
          }

   \institute{
   	Univ. Grenoble Alpes, CNRS, IPAG, 38000 Grenoble, France 
	\and
	Department of Physics, University of Durham, South Road, Durham, DH1 3LE, UK
	\and
	Department of Astronomy, University of Michigan, 1085 South University Avenue, Ann Arbor, MI 48109, USA
	\and
	Universit\'e de Strasbourg, CNRS, Observatoire Astronomique de Strasbourg, UMR 7550, 67000 Strasbourg, France
	     }

   \date{Received ; accepted ; in original form \today}

  \abstract
	{The observed signatures of winds from X-ray binaries are broadly consistent with thermal winds, driven by X-ray irradiation of the outer accretion disc. Thermal winds produce mass outflow rates that can exceed the accretion rate in the disc.}
   {We aim to study the impact of thermal wind mass loss on the stability and lightcurves of black hole X-ray binaries subject to the thermal-viscous instability, which drives their outbursts. Strong mass loss could shut off outbursts early, as proposed for the 2015 outburst of V404 Cyg.}
   {We use an analytical model for thermal (Compton) wind mass loss as a function of radius, X-ray spectrum and luminosity that has been calibrated against numerical simulations. We also estimate the fraction of the X-rays, emitted close to the compact object, that are scattered back to the outer disc in the wind. Scattering in the thermal wind couples irradiation to the disc size and inner mass accretion rate. The disc evolution equation are modified to include this wind mass loss and varying irradiation fraction.}
   {Scattering in the strong wind expected of long $P_{\rm orb}$ systems enhances the irradiation heating of the outer disc, keeping it stable against the thermal-viscous instability. This accounts very well for the existence of persistently bright systems with large discs such as Cyg X-2, 1E 1740.7$-$2942, or GRS 1758$-$258. Including mass loss from the thermal wind shortens the outburst, as expected, but insufficiently to explain the rapid decay timescale of black hole X-ray binary outbursts. However, including the wind-related varying irradiation fraction produces lightcurves with plateaus in long $P_{\rm orb}$ systems like GRO J1655$-$40. Plateau lightcurves may be a dynamical signature of enhanced irradiation due to scattering in a thermal wind.}
  	{Mass loss due to thermal winds is not a major driver for the outburst dynamics up to luminosities $0.1-0.2 L_{\rm Edd}$. Higher luminosities may produce stronger mass loss but their study is complicated since the wind becomes opaque. Magnetic winds, which extract angular momentum with little mass loss, seem more promising to explain the fast decay timescales generically seen in black hole X-ray binaries. Thermal winds can play an important role in the outburst dynamics through the varying irradiation heating. This may be evidenced by relating changes in wind properties, X-ray spectra or luminosity, with changes in the optical emission that traces the outer disc. Simulations should enable more accurate estimates of the dependence of the irradiation onto the disc as a function of irradiation spectrum, radius and disc wind properties.}

   \maketitle
%

\section{Introduction}

Recurring outbursts due to the hydrogen ionisation disc instability provide test beds for constraining key aspects of accretion processes \citep{Lasota:2001th}. These are seen from compact objects of all types
but the  black-hole X-ray binaries (BHXBs), are of particular interest.
Their bright X-ray and optical outbursts typically last hundreds of days and recur on timescales of years \citep{tetarenko2016}. This provides an opportunity to study accretion over astrophysically interesting and observable timescales. 

The large change in mass accretion rate through the disc produces a distinct pattern of spectral change,
with the system cycling between hard (dominated by Comptonized emission) and soft (dominated by thermal disc emission) accretion states. This is generally interpreted as a change in the nature of the accretion flow, from a hot, optically thin, geometrically thick solution (such as an Advection Dominated Accretion Flow: ADAF) to a
cool, geometrically thin standard disc \citep{mcclintock2006,done2007}. 

X-ray irradiation from either the inner disc or the hot inner flow can illuminate the outer disc. This makes a major contribution to the thermal balance at large radii, so the optical emission is dominated by 
reprocessed X-rays \citep{vanparadijs1994,van-Paradijs:1996dz}. This irradiation heating changes the dynamics of the transient outburst, as it prevents hydrogen recombining, keeping the outer disc in the hot state and prolonging the outburst timescale \citep{1998MNRAS.293L..42K,2001A&A...373..251D}. 

The accretion energy also powers two different types of outflows which change with accretion state. First, there is a compact, steady jet. This jet is associated with the hot flow, so the collapse of the flow at the spectral state transition leads to strong suppression of the radio jet emission, though it can restart at higher luminosities, again correlated with the strength of the X-ray corona (very high/steep power law state). Second, there are strong equatorial disc winds which are associated with the disc dominated states, but seem to disappear in the hard state \citep{miller2006,ponti2012,neilsen2013,diaztrigo2014}. 
There are a number of possible explanations for suppression of the wind features in the hard state including  (i) a causal connection, with the same magnetic field re-configuring to power a jet in the hard state and a wind in the soft state \citep{neilsen2009,miller2012}, and (ii) correlation, where changing spectral shape during state transitions results in complete ionisation of the wind material in the hard accretion state \citep{chakravorty2013,higginbottom2015}. The former mechanism assumes that the wind is magnetically launched, in which case it can also contribute to or even completely control the angular momentum transport in the accretion flow \citep{1982MNRAS.199..883B,1995A&A...295..807F,scepi2018}. All winds will change the disc instability timescales by removing mass from the disc, but magnetic winds should give an additional change in the  disc instability timescales by enhancing the angular momentum transport as well \citep{2019A&A...626A.116S}. 

However, there are other wind launch mechanisms, such as radiative or thermal driving, which remove matter from the flow without torquing the disc. Radiative driving becomes important only for $L>L_{\rm Edd}$ (assuming no additional opacity beyond electron scattering), but thermal driving inevitably produces ``Compton'' winds from irradiation of the outer disc. X-ray heating by irradiation from the inner disc forms a dilute, isothermal corona at the Compton temperature $T_{\rm IC}$ set by the spectral shape of the incident radiation. This also sets a characteristic radius, the Compton radius $R_{\rm IC}$, at which the sound speed of this material becomes greater than its escape velocity, leading to an outflow 
 \citep{1983ApJ...271...70B}.  Numerical simulations of these thermal winds have mostly confirmed this picture, with the most recent simulations taking into account detailed radiative transfer, cooling function and ionisation balance \citep{1996ApJ...461..767W,2002ApJ...565..455P,2010ApJ...719..515L,2019MNRAS.484.4635H}.
 
While some of the observed velocities, ionisation fractions and column densities in disc winds are broadly consistent with thermal winds \citep{2016AN....337..368D}, there are some much debated exceptions.
The seminal detection of a strong wind in GRO J1655-40 was used to argue strongly for magnetic winds as it 
appeared  impossible to produce by any alternative mechanism \citep{miller2006}.
Radiative driving could be ruled out as the source was observed to be highly sub-Eddington at $L\sim 0.05L_{\rm Edd}$, and the inferred launch radius determined directly from a density diagnostic gave $R\ll 0.1 R_{IC}$.
However, multiple authors now suggest that the wind has gone optically thick, so that the observed luminosity is strongly underestimated and is instead close to (or above) Eddington (see \citealt{2018MNRAS.481.2628W} for a discussion of this source), and it is now known that the metastable transitions can be populated by UV pumping as well as by collisional excitation, so they are not a clean density diagnostic. However, there is also other evidence for more complex winds. Photo ionization modelling of higher-resolution (third-order) Chandra grating spectra provides the ability to mostly resolve absorption line profiles into multiple components. Using this technique,  \citet{miller2015,miller2016} have reanalyzed Chandra spectra on the four known BHXB wind sources, and claim some components require small launching radii, high density values, and inferred high mass outflow rates. They suggest that there may be magnetic winds in addition to the thermal winds, or magneto-thermal winds  (see also \citealt{chakravorty2016}). 

The jet-wind connection is even more unclear. The initial anti-correlation of wind and jet \citep{neilsen2009,ponti2012} has now been replaced with a more complex picture, where jets and winds are seen together in the very high/steep power law state \citep{homan2016} and optical/IR winds are detected via P\,Cygni line profiles in H$\alpha$ and HeI in the hard state (e.g., V4641 Sgr and MAXI J1820+070; \citealt{munozdarias2018,munozdarias2019}).

Thus, currently, the principle driving mechanism for these disc winds inferred by observational X-ray and optical/IR spectra, whether it be thermally-driven (by X-ray irradiation) or magnetically-driven (by magnetic pressure and centrifugal acceleration along magnetic field lines anchored in the disc), remains a matter of great debate. Instead, we take an alternative approach, and investigate the wind launching mechanism(s) by exploring their effect on the disc instability lightcurves.

\cite{2018Natur.554...69T}, developed a methodology to effectively characterize the angular momentum transport processes in BHXB accretion discs from their X-rays lightcurves. This gave quantitative measurements of the $\alpha$ viscosity \citep{shakura1973}, the parameter which encapsulates the efficiency of this transport process, in these discs. While numerical simulations of the magneto-rotational instability \citep{balbus1998}, which is thought to be the physical mechanism behind angular momentum transport in accretion discs, yield values of $0.1<\alpha<0.2$ with zero net magnetic flux \citep{hirose2014,coleman2016,scepi2018}, \cite{2018Natur.554...69T} derive values of $\alpha$ that are significantly higher, between $0.2-1$. As $\alpha$ is directly related to the viscous timescale of the accretion flow through the disc, a higher $\alpha$ during outburst corresponds to a disc that will accrete mass more quickly, resulting in a shorter outburst duration then would normally occur. \cite{2018Natur.554...69T} propose that the high values of $\alpha$ derived, and in-turn shorter than expected outburst durations observed, in BHXB discs, may be evidence for significant amounts of mass (and/or angular momentum) being lost from the disc through wind type outflows during these outbursts \citep{1993PASJ...45..707M}.

We explore the mass loss possibility by incorporating thermal winds into the disc instability code. We adopt the wind parametrisation described by \citet{2018MNRAS.473..838D}, which gives us the wind mass loss rate in the disc as a function of luminosity. This parametrisation typically predicts that the wind outflow rate $\dot{M}_{\rm w}$ can be several times the accretion rate through the disc, when the disc is large compared to the Compton radius, in agreement with the latest simulations \citep{2019MNRAS.484.4635H,2019arXiv190511763T} and observations \citep{2019arXiv190602469S}. We use this wind to scatter the central flux down onto the disc, providing a physical quantitative estimate for the size of this effect to replace the previous ad hoc irradiation parameter (i.e., ${\cal C}$; \citealt{dubus1999}). The wind strength increases with disc size and luminosity, so the scattered fraction is not constant, but changes during the outburst. We incorporate this effect in addition to the wind mass loss, and calculate the resulting disc instability lightcurves. We find notable differences in the outburst timescales, including the first calculations showing extended plateaus in the lightcurves, a feature often seen in the data, but not reproduced by any previous models. The models do not currently reproduce the apparent high $\alpha$ viscosity values, which could indicate that there is an additional magnetic angular momentum transport mechanism. However, we caution that these simulated outbursts currently only reach moderate luminosities of $0.1-0.2L_{\rm Edd}$, so they are not yet directly comparable to many of the observed systems \citep{yan2015,tetarenko2016}. 

The paper is organized as follows. Section 2 provides a description of the model. Section 3 explains how thermal winds change the disc structure and stability. Section 4 presents our results for one short period and one long period system. Section 5 discusses these selected models of outbursting BHXBs. We conclude in Sect. 6.

\section{The accretion disc model\label{s:model}}

\subsection{The disc equations}
In the presence of mass loss from a wind, the continuity equation is rewritten with an additional loss term
\begin{equation}
\pd{\Sigma }{t}-\frac{1}{2\pi r} \pd{\dot{M}}{r}=-\dot{\Sigma}_{\rm w}
\end{equation}
and the specific angular momentum conservation equation becomes
\begin{equation}
\Sigma \pd{r^2 \Omega}{t}-\frac{\dot{M}}{2\pi r}\pd{r^2\Omega}{r} = \frac{1}{r}\pd{}{r}\left(r^3\nu\Sigma\pd{\Omega}{r} \right)-j\dot{\Sigma}_{\rm w}
\label{eq:eqmo}
\end{equation}
where $j$ is the excess specific angular momentum carried away by the wind \citep{1995ARA&A..33..505P}. Unlike magnetic winds, thermal winds are not expected to drive excess angular momentum loss so we took $j=0$ \citep{1999MNRAS.309..409K}. We also assume a Keplerian rotation profile, as appropriate for thin accretion discs, so the first term of Eq.~\ref{eq:eqmo} drops out. Hence, the only change to \citet{2001A&A...373..251D} is the continuity equation.
We define the wind mass loss rate at a radius $R$ as
\begin{equation}
\dot{M}_{\rm w}(R)\equiv \int_{R}^{R_{\rm out}} 2\pi r \dot{\Sigma}_{\rm w} dr
\end{equation}
so that the total mass outflow rate in the wind is $\dot{M}_w(R_{\rm in})$.

\subsection{The disc wind}
The $\dot{\Sigma}_{\rm w}$ term due to the thermal wind is the $\dot{m}$ given by Eq. 4.8 from \citet{1996ApJ...461..767W}, using the assumptions appropriate to X-ray binaries set out in \citet{2018MNRAS.473..838D}. We refer to these papers for the complete expression of this term. The wind is all the more important that the disc is large and the luminosity is high. The X-ray irradiation luminosity is taken to be $L=\epsilon \dot{M}_{\rm in} c^2$ with the radiative efficiency $\epsilon$ as defined in \citet{2001A&A...373..251D}, and $\dot{M}_{\rm in}\equiv\dot{M}(R_{\rm in})$.

The thermal wind is effective for radii $R\ga 0.2 R_{\rm IC}$, with the Compton radius $R_{\rm IC}$ set as
\begin{equation}
R_{\rm IC}\approx 10^{12} \left(\frac{M}{10 M_\odot}\right) \left(\frac{10^7\rm\,K}{T_{\rm IC}}\right)\left(1-\sqrt{2}\frac{L}{L_{\rm Edd}}\right)\rm\,cm
\label{eq:ric}
\end{equation}
where $M$ is the compact object mass, $T_{\rm IC}$ is the Compton temperature of the impinging radiation, and we include a rough correction to take into account radiation pressure on electrons at luminosities close to Eddington \citep{2002ApJ...565..455P,2018MNRAS.473..838D}.

We followed the assumptions set out in \citet{2018MNRAS.473..838D}. In particular, a switch in Compton temperature $T_{\rm IC}$ occurs between hard and soft state due to the major change in X-ray irradiation spectrum.  The state transition is taken to occur at a fixed luminosity $l\equiv L/L_{\rm Edd}=0.02$ with (Eqs.~8-9 of  \citealt{2018MNRAS.473..838D}) 
\begin{equation}
       \frac{T_{\rm IC}}{10^7\rm\,K} = 
        \begin{cases}
            4.2 - 4.6\log\left(\frac{l}{0.02}\right) & \text{if $l<0.02$ (hard state)} \\
            0.36 \left(\frac{l}{0.02}\right)^{1/4} & \text{if $l\geq 0.02$ (soft state)} \\
        \end{cases}
        \label{eq:tic}
\end{equation}
In the hard state, $T_{\rm IC}$ decreases slowly from $\approx 10^8\rm\,K$  at $l=10^{-4}$ to about $4\times 10^7\rm\,K$ at $l=0.02$, where it abruptly switches to the soft state with $T_{\rm IC}\approx 4\times 10^6\rm\,K$, increasing slowly to $10^7\rm\,K$ at $l=1$ (see top panel of Fig.~5 in \citealt{2018MNRAS.473..838D}). To avoid the discontinuity at $l=0.02$, we smoothly switch $T_{\rm IC}$ according to
\begin{equation}
T_{\rm IC}=T_{\rm HS} \left(\frac{1}{1+x}\right) + T_{\rm SS} \left(\frac{x}{1+x}\right)~\text{with}~x=\left(\frac{l}{0.02}\right)^6.
\label{eq:trans}
\end{equation}
$T_{\rm HS}$ and $T_{\rm SS}$ are the Compton temperature in the hard and soft state, respectively. We verified that this choice, which eases the computations, is of no consequence to the lightcurves.

A thermal wind is more easily produced in the hard state than in the soft state, due to the smaller  $R_{\rm IC}$ (higher  $T_{\rm IC}$), and can be quenched during the transition to the soft state due to the increase in $R_{\rm IC}$. However, the ratio of the mass loss rate in the wind to the mass accretion rate in the inner disc also increases with luminosity up to the critical luminosity,
\begin{equation}
L_{\rm crit}\approx 0.09 \ \left(\frac{10^7\rm\,K}{T_{\rm IC}}\right)^{1/2} L_{\rm Edd}.
\end{equation}
This critical luminosity is set by the balance between the Compton heating timescale and the dynamical timescale of the disc at the Compton radius \citep{1996ApJ...461..767W}. In practice, the outflow rate in the wind is about 2 to 5 times the mass inflow rate in the disc for $l\ga 0.001$  (bottom panel of Fig.~5 in \citealt{2018MNRAS.473..838D}). We expect thermal wind mass loss to be dynamically important for discs with an outer radius $R_{\rm out}\ga 2\times 10^{11} \rm\,cm$ for a 10$\,M_\odot$ black hole, i.e. for long $P_{\rm orb}$ binaries. We do not take into account that the forward transition (from the hard to  soft state) occurs at a higher luminosity during the outburst rise than the backwards transition (from the soft to hard state) i.e. hardness -- intensity hysteresis.

\subsection{Irradiation heating}
The irradiation flux heating the accretion disc at a radius $R$ is defined as
\begin{equation}
F_{\rm irr}\equiv\sigma T_{\rm irr}^4\equiv{\cal C}\frac{ L}{4\pi R^2}
\end{equation}
where ${\cal C}$ encapsulates uncertainties in the geometry and albedo of the outer disc irradiation by X-rays \citep{1990ApJ...359..164T,dubus1999}. Typically, ${\cal C}$ is assumed to be a constant of value $\approx 5\times 10^{-3}$ (see \citealt{2018MNRAS.480....2T} and references therein). For comparison, \citet{2019MNRAS.482..626K} considered direct irradiation of an isothermal disc by a central point source
\begin{equation}
{\cal C}=(1-A) \frac{H}{R} \left(\frac{{\rm d}\ln H}{{\rm d}\ln R}-1\right) \approx 10^{-3} \left(\frac{R}{R_{\rm out}}\right)^{2/7}
\end{equation}
assuming $H/R=f_{\rm out} (R/R_{\rm out})^{2/7}$ with $f_{\rm out}=4\times 10^{-2}$ and $(1-A)=0.1$  \citep{1976ApJ...208..534C, 1990A&A...235..162V}. This results in ${\cal C}\propto R^{2/7}$ while a standard Shakura Sunyaev disc would give ${\cal C}\propto R^{1/8}$ \citep{1976ApJ...208..534C}, hence the dependency of $\cal C$ with radius is weak in both cases. We note that $L=\epsilon \dot{M}_{\rm in} c^2$ with $\epsilon$ changing throughout the outburst to reflect the reduced radiative efficiency when an ADAF is present (see \citealt{2001A&A...373..251D} for details).

In principle, the flux irradiating the outer disc and the conditions for launching a thermal wind are related not only by $L$ but also by the underlying irradiation geometry (the $\cal C$ parameter). The thermal wind may itself provide a medium to scatter X-ray photons to the disc. This is attractive since point source irradiation is known to be insufficient to explain the stability of persistent X-ray binaries: the cold, outer region that should be irradiated has a smaller scale height than the hotter, inner region and is thus shadowed from the X-ray source \citep{1982A&A...106...34M,cannizzo1995,dubus1999,kim1999}. The irradiation geometry must allow the X-ray flux to irradiate the outer disc e.g., because the X-ray source is a ``lamp post'' above the disc, because the disc is warped, or because the X-ray flux is scattered in a corona or wind above the disc. \citet{2019MNRAS.482..626K} estimated the scattered fraction in the thermal wind to be comparable to or higher than $f_{\rm out}$, in agreement with the earlier results of \citet{Ostriker1991} who studied the scattered fraction in a Compton heated corona (i.e. when $R\la 0.1 R_{\rm IC}$) and found that it dominates over direct irradiation when $L\ga 0.01 L_{\rm Edd}$ (see their section 7.4). We explored this idea by having $\cal C$ vary according to 
\begin{equation}
{\cal C}=\int_0^1 \int_{R_{\rm in}}^{R_{\rm out}} \sigma_T n_{\rm w} \mu d\mu dr \approx \frac{\sigma_T \dot{M}_{\rm w}}{8\pi R_{\rm in} v_{\rm w} m_I} \propto L  \log \left(\frac{R_{\rm out}}{R_{\rm in}}\right)
\label{eq:estC}
\end{equation}
with $n_{\rm w}$ the wind density, $v_{\rm w}$ the mass-weighted wind outflow rate, $m_I$ is the mean ion mass per electron and $\mu=\cos i$ (see Eq.\,6 in \citealt{2018MNRAS.473..838D}, and the wind property dependencies given in \citealt{Hori2018}). 
$\cal C$ is basically the angle-averaged Thomson opacity of the wind, which is $\ll 1$ in most cases investigated here. All the major dependencies on $T_{\rm IC}$ cancel out, so the scattered fraction correlates directly with $L$, with the Compton temperature only indirectly contributing via $R_{\rm in}=0.2R_{\rm IC}$ in the logarithm.

\section{Impact of thermal winds on the stability of X-ray binaries}
\subsection{The radial structure of a stationary disc with a thermal wind}
\begin{figure}
\begin{center}
\includegraphics[width=0.8\linewidth]{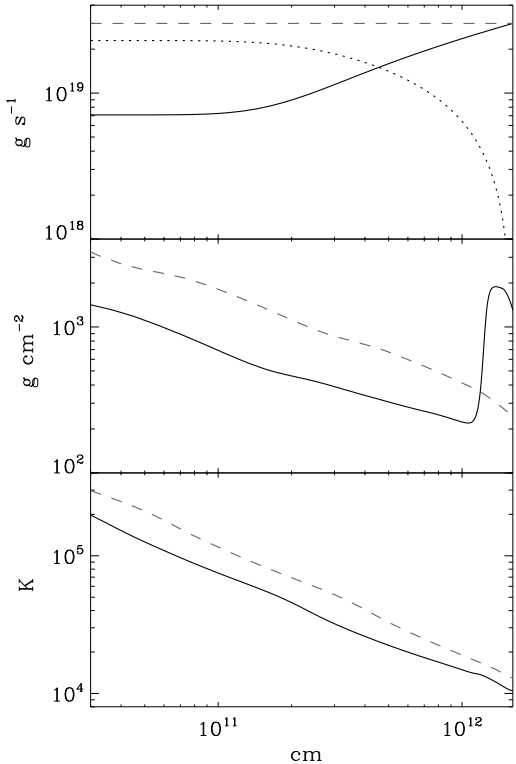} 
\caption{Radial structures for a disc around a 12\,M$_\odot$ black hole with (full lines) and without (dashed lines) a thermal wind. Top panel shows $\dot{M}$ and $\dot{M}_{\rm w}$ (dotted line, see Eq.~\ref{eq:mdotwind}). Middle panel shows $\Sigma$. Bottom panel shows the central temperature $T_{\rm c}$. }
\label{fig:stat}
\end{center}
\end{figure}
Figure \ref{fig:stat} compares the radial structure of stationary solutions ($\partial/\partial t=0$) with and without a thermal wind. The disc size $R_{\rm out} \approx 1.6\times 10^{12}\rm\,cm$ and the mass transfer rate $\dot{M}_{\rm t}=3\times 10^{19}\rm\,g\,s^{-1}$ are large so as to maximise the effect of the thermal wind. The compact object is a 12\,M$_\odot$ black hole so the orbital period of the binary would be  around 10 days i.e. this is a large system. The accretion luminosity is slightly super-Eddington without a thermal wind (dashed lines in Fig.\,\ref{fig:stat}). The irradiation constant is ${\cal C}=0.01$ so the outer disc is hot and stable. This value of ${\cal C}$ was chosen to be close to the value derived from scattering in the wind (${\cal C}\approx 0.05$ using Eq.\,\ref{eq:estC}) while still allowing for the outer disc to be unstable when a wind is included, as we now explain.

With a thermal wind, the mass accretion rate through the disc becomes a function of $R$ with,  
\begin{equation}
\dot{M}_{\rm t}-\dot{M}(R)=\dot{M}_{\rm w}(R).
\label{eq:mdotwind}
\end{equation}
The total mass lost to the wind over the whole disc is $\dot{M}_{\rm w}(R_{\rm in})\approx 2.3 \times 10^{19}\rm\,g\,s^{-1}$. The top panel shows that $\dot{M}(R)\sim R^{1/2}$ in the outer regions, becoming constant where $\dot{\Sigma}_{\rm w}=0$ i.e. for $R<0.2 R_{\rm IC}$ ($R\la 1.6\times 10^{11}\rm\,cm$ here). The lower $\dot{M}_{\rm in}$ decreases the luminosity to $\approx 0.35 L_{\rm Edd}$. The outer regions now reach the critical temperature for hydrogen recombination, signalled by a jump in surface density $\Sigma$ (the solution is forced to follow the unstable middle branch of the S curve). Another way to see this is to note that Eq.~\ref{eq:eqmo} can be rewritten,
\begin{equation}
    \pd{}{r}\left(\nu \Sigma r^{1/2}\right)=\frac{\dot{M}}{6\pi} r^{-1/2}\approx \frac{\dot{M}_{\rm t}}{6\pi} R_{\rm out}^{-1/2},
\end{equation}
given the scaling with radius of $\dot{M}$. This implies that $\nu \Sigma\approx \dot{M}/6\pi$, a factor 2 lower than in a standard disc (where $\dot{M}$ would be constant with radius). Since $Q^+\propto \nu \Sigma$, the heating is decreased by a factor 2. Thus, the thermal wind destabilizes an accretion disc that would otherwise be stable. However, increasing $\cal C$ to 0.04 would be sufficient to stabilize the disc again. In fact, ${\cal C}\approx 0.05$ when calculated from the scattering in the wind (Eq.\,\ref{eq:estC}). A self-consistent determination would thus have found the disc remained stable even with a thermal wind, as we now explain.

\subsection{The stability diagram\label{sec:stab}}

The disc instability model separates stable X-ray binaries from transients in the $(P_{\rm orb},\dot{M}_{\rm t})$ plane \citep{van-Paradijs:1996dz,2012MNRAS.424.1991C,tetarenko2016}. Thermal winds can have three effects. First, as demonstrated above, they can destabilise long $P_{\rm orb}$ systems accreting at very high rates. Second, they lead to an underestimate of the mass transfer rate $\dot{M}_{\rm t}$ if  the mean X-ray luminosity is used to estimate it. For example, the mass transfer rate would be underestimated by a factor $\dot{M}(R_{\rm out})/\dot{M}(R_{\rm in})\approx 4$ for the disc shown in Fig.~\ref{fig:stat}. \citet{2012MNRAS.424.1991C} proposed this to explain why Cyg X-2 is stable although its estimated $\dot{M}_t$ places it in the unstable region of the instability diagram. Finally, the value of $\cal C$ is not constant anymore with disc size ($P_{\rm orb}$) and mass transfer rate but will depend on the wind properties.

\begin{figure*}
\begin{center}
\includegraphics[width=0.45\linewidth]{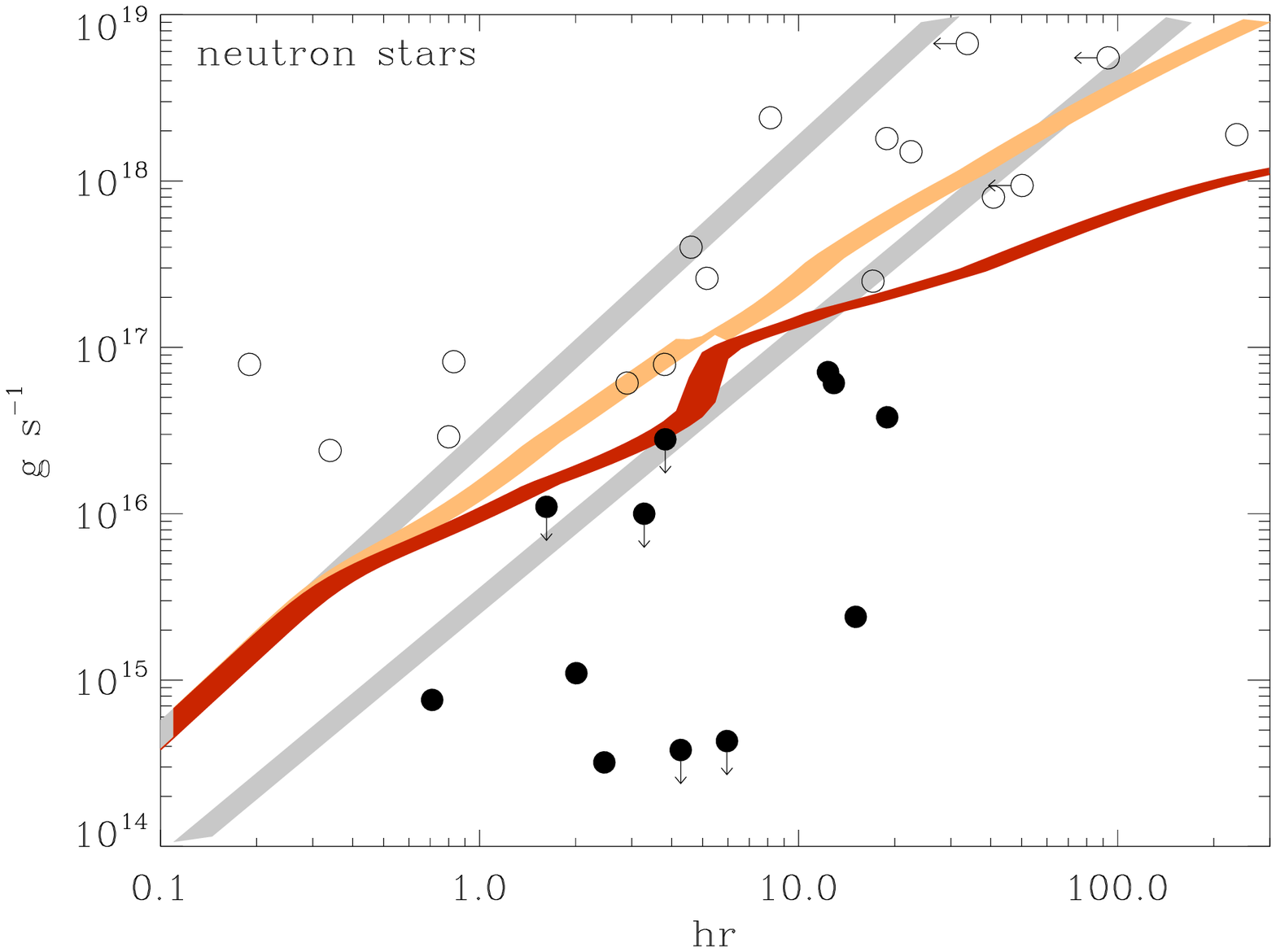} 
\includegraphics[width=0.45\linewidth]{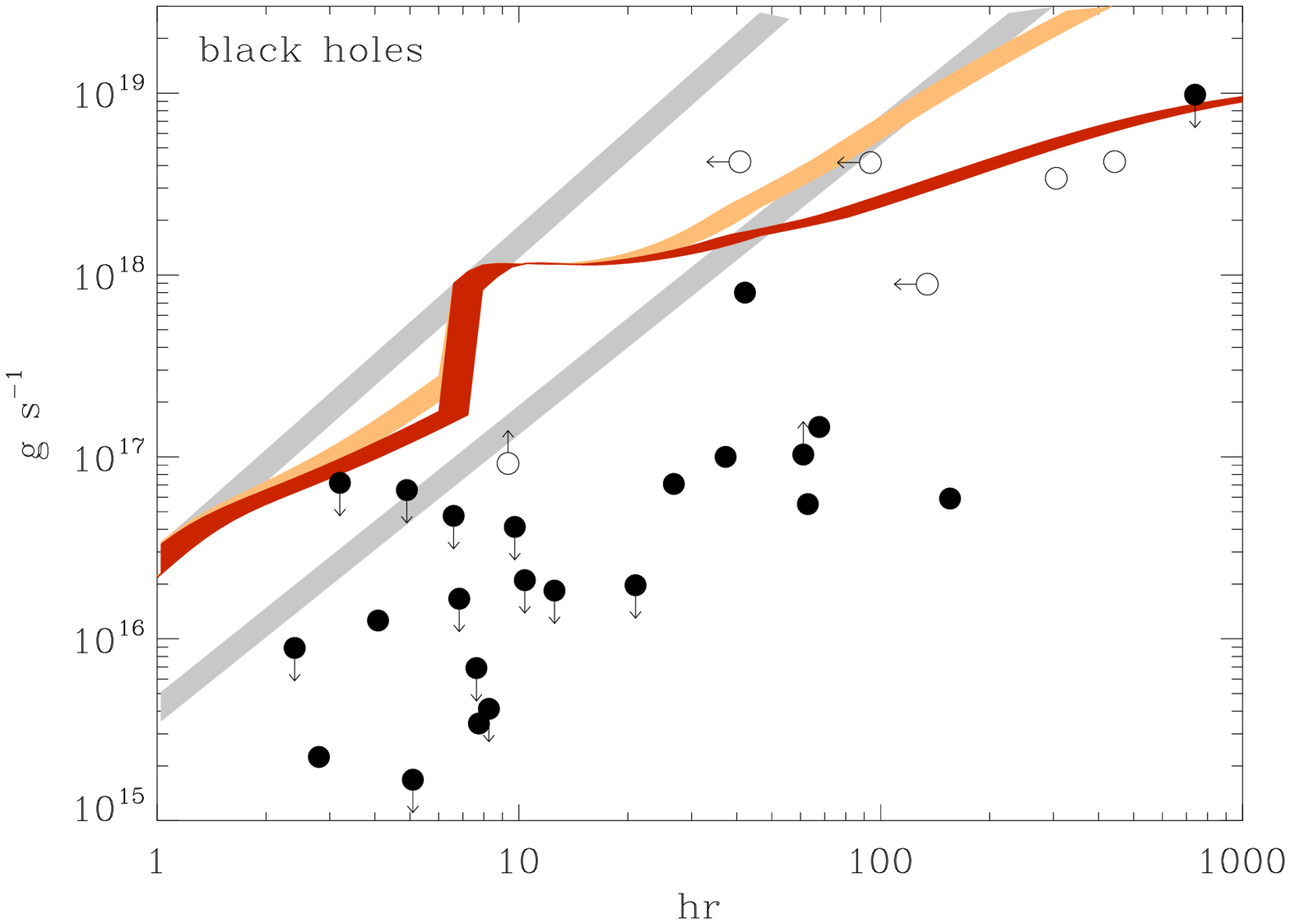} 
\caption{Stability diagram for neutron stars (left panel) and black holes (right panel) in the ($P_{\rm orb}$, $\dot{M}$) plane. Data points taken from \citet{2012MNRAS.424.1991C} and the WATCHDOG catalogue \citep{tetarenko2016}, with open and filled circles representing stable and unstable systems, respectively. In each panel, the two straight grey lines show the stability limits for the standard unirradiated (top line) and irradiated (bottom line, with ${\cal C}=10^{-3}$) discs. The red line shows the stability limit on $\dot{M}_{\rm in}$ with a thermal wind and a self-consistent determination of $\cal C$. The yellow line shows the same but on the mass transfer rate $\dot{M}_{\rm t}=\dot{M}_{\rm w}+\dot{M}_{\rm in}$. The limits are calculated for a $1.4\rm\,M_\odot$ neutron star and a $10\rm\,M_\odot$ black hole. The thickness of the lines illustrates how the stability criterion changes when the mass ratio of the secondary star to the compact object $q$ changes from 0.1 to 1.}
\label{fig:stab}
\end{center}
\end{figure*}

Figure \ref{fig:stab} shows the impact on the stability curve of thermal winds, taking into account all these effects. The usual stability criterions are shown as grey lines. For the non-irradiated case, this is,
\begin{equation}
\dot{M}^{\rm non\,irr}_{\rm crit}\ga 8.07\times 10^{15} R_{10}^{2.64} M_1^{-0.89}\rm\,g\,s^{-1},
\end{equation}
with $M=M_1 \rm\,M_\odot$ the mass of the compact object and $R_{\rm out}=R_{10} \times 10^{10}\rm\,cm$; in the irradiated case, the criterion is,
\begin{equation}
\dot{M}^{\rm irr}_{\rm crit}\ga 9.5\times 10^{14}\ {\cal C}_{3}^{-0.36}  R_{10}^{2.39-0.1\log{\cal C}_{3}}M_1^{-0.64+0.08\log{\cal C}_{3}}\rm\,g\,s^{-1},
\label{eq:irr}
\end{equation}
with ${\cal C}=10^{-3}{\cal C}_{3}$. Both parametrisations are taken from  Appendix A of \citet{2008A&A...486..523L}. The line thickness for a given $P_{\rm orb}$ in Fig.\,\ref{fig:stab} corresponds to the range in $R_{\rm out}$ for $0.1\leq q\leq 1$, with $q$ being the mass ratio between secondary star and compact object.

Equation \ref{eq:irr} becomes a non-linear equation to be solved when $\cal C$ is taken from the scattered fraction i.e., when it becomes dependent on  $R_{\rm out}$ and accretion rate $\dot{M}_{\rm in}$. A self-consistent solution gives the $\dot{M}_{\rm in}$ that provides the right $\cal C$ for the disc to be stable at $\dot{M}_{\rm in}$ when irradiated by a luminosity $L=\epsilon \dot{M}_{\rm in}c^2$. Here, we assumed $\epsilon=0.1$. We took the solution with the lowest $\dot{M}_{\rm in}$ when two solutions to the equation were possible, one in the soft state and one in the hard state. Two solutions exist in a narrow region close to the transition luminosity. This could lead to a situation in which a system is stable with the hard state $T_{\rm ic}$ yet unstable if it switches to the soft state $T_{\rm ic}$, due to a weaker wind and $\cal C$. We have not investigated this further as it depends on the assumed form of the transition between states (Eq.\,\ref{eq:trans}).

The red line shows the resulting self-consistent $\dot{M}_{\rm in}$. The yellow line shows $\dot{M}_{\rm t}=\dot{M}_{\rm in}+\dot{M}_{\rm w}$, showing that the mass transfer rate from the companion is much higher than the accretion rate at long $P_{\rm orb}$. The $\dot{M}_{\rm in}$ stability line is flatter than in the case with no wind and constant $\cal C$. This is because $\cal C$ increases with longer $P_{\rm orb}$ as the disc becomes larger and the wind stronger. The stability curve joins the non-irradiated standard one at short $P_{\rm orb}$ because the wind and, correspondingly, $\cal C$ are weak for small discs. The kink or jump in the curve at $P_{\rm orb}\approx 7\rm\,hr$ is due to the hard to soft switch at $L=0.02 L_{\rm Edd}$. This jump is stronger for black holes than neutron stars because it occurs at a higher absolute luminosity whereas the disc sizes remain similar. The location of the jump and its amplitude depend on the assumptions made on $T_{\rm IC}$ and the transition luminosity between states (Eq.\,\ref{eq:tic}). In particular, we use the same assumptions for black holes and neutron stars, which may not be appropriate.  A higher $T_{\rm IC}$ would move the jump to longer $P_{\rm orb}$ and reduce its amplitude. 

The data for the different systems is taken from \citet{2012MNRAS.424.1991C}, supplemented by data from the WATCHDOG catalogue for 6 transient black hole systems: MAXI J1305$-$704, SWIFT J1357.2$-$0933, V4641 Sgr, MAXI J1836$-$194, SWIFT J174510.8$-$262411, XTE J1859$+$226 \citep{tetarenko2016}. They calculated the time-averaged average mass accretion rate of each system from their long-term X-ray lightcurves with some assumptions on $\epsilon$. These values thus do provide an average $\dot{M}_{\rm in}$ to compare to the red line. 

Figure \ref{fig:stab} shows that the stability line separates extremely well persistent systems from transients with no free parameters. The agreement is improved for long $P_{\rm orb}$ sources. Indeed, Cyg X-2 (at $P_{\rm orb}=236\rm\,hr$ in the left panel) is now above the stability line for neutron stars. For black holes, 1E 1740.7$-$2942 ($P_{\rm orb}=305\rm\,hr$) and GRS 1758$-$258 $P_{\rm orb}=443\rm\,hr$) are two persistent or long duration outburst sources that are now very close to the stability line. However, we strongly caution that the derived $\cal C$ become unphysical for long $P_{\rm orb}$ since the wind becomes Compton thick: ${\cal C}\geq 0.1$ for $P_{\rm orb}\ga 150\rm\,hr$ (1.4\,M$_\odot$ neutron star) and for $P_{\rm orb}\ga 500\rm\,hr$ (10\,M$_\odot$ black hole). 

There are a few remaining discrepancies for BHXBs (see discussion in \citealt{2012MNRAS.424.1991C}). 4U 1957$+$115 at $P_{\rm orb}=9.3\rm\,hr$ is persistent and in the unstable region, close to the jump caused by the change in $T_{\rm IC}$. The lower limit on its $\dot{M}$ comes from a lower limit on its distance. Cyg X-1 at $P_{\rm orb}=134\rm\,hr$ is also a persistent system in the unsteady region of the diagram. The leftward arrow on its symbol indicates the actual disc size of this wind-accreting system could be much smaller than the values derived from Roche lobe mass transfer. In both cases, a calculation with the  observed $T_{\rm IC}$  for this system, rather than generic values, might reduce the discrepancy by moving the red line rightward.

\section{Impact of mass loss on outburst lightcurves}
\subsection{A short orbital period system\label{sporb}}
\begin{figure}
\begin{center}
\includegraphics[width=\linewidth]{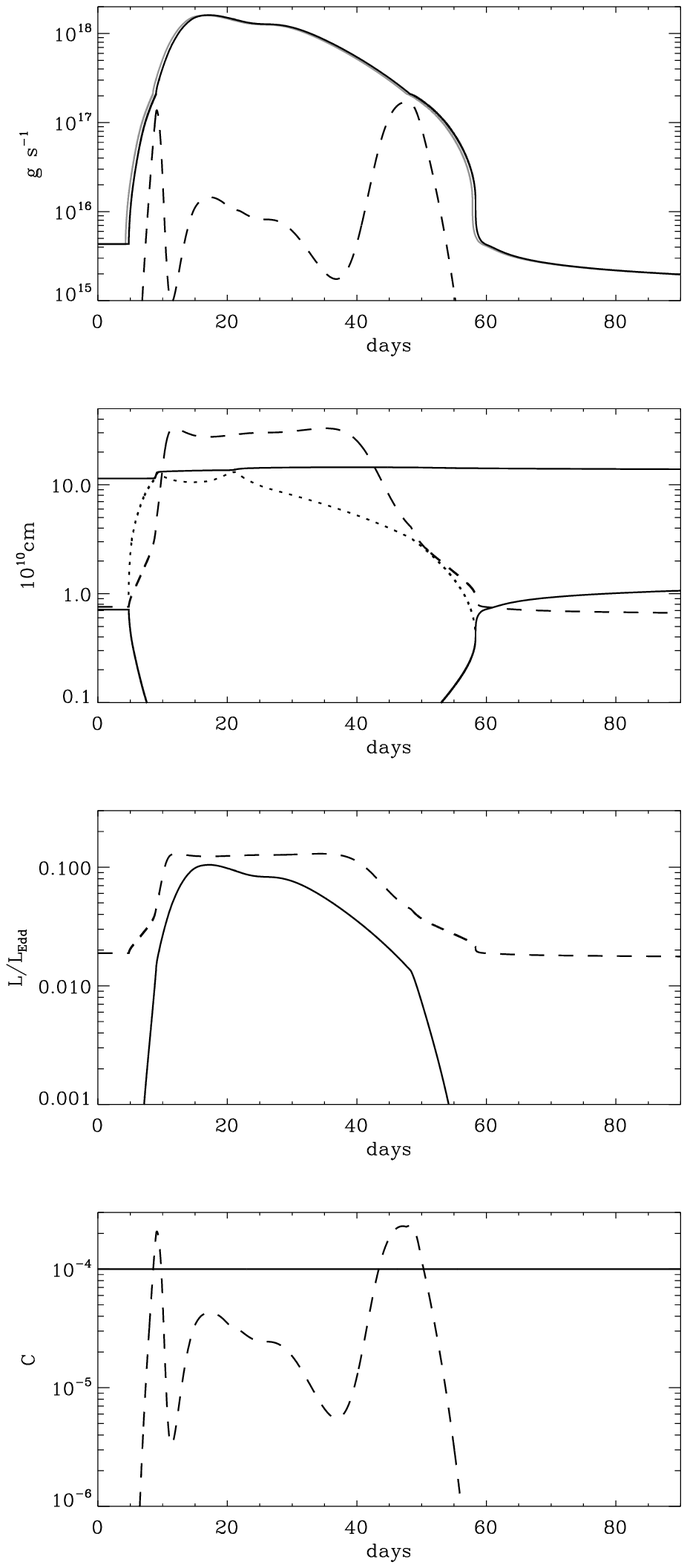} 
\caption{Outburst for a short orbital period black hole. Top panel: $\dot{M}(R_{\rm in})$ (black solid line), $\dot{M}_w(R_{\rm in})$ (dashed line),  $\dot{M}(R_{\rm in})$ no wind mass loss for comparison (grey solid line, nearly superposed to the black solid line). Second panel: $R_{\rm in}$ and $R_{\rm out}$ (solid lines), radius of the heating/cooling front (dotted line), $0.2 R_{\rm ic}$ (dashed line). Third panel: $L/L_{\rm Edd}$ (solid line) and $L_{\rm crit}/L_{\rm Edd}$ (dashed line). Bottom panel: $\cal C$ (solid line) and $\cal C$ estimated from Eq.\,\ref{eq:estC} (dashed line).}
\label{fig:206a}
\end{center}
\end{figure}

First, we consider an unstable accretion disc around a $M=9\rm\,M_\odot$ black hole with a mass transfer rate $\dot{M}_{\rm t}=10^{16}\rm\,g\,s^{-1}$. The disc has a size $\approx 10^{11}\rm\,cm$, corresponding to an orbital period $\approx 5\rm\,hr$. The disc has a viscosity parameter $\alpha_{\rm h}=0.2$ in the hot state and $\alpha_{\rm c}=0.02$ in the cold state, typical values that have been used in previous disk instability model calculations \citep{2001A&A...373..251D}. The irradiation parameter is assumed constant at ${\cal C}=10^{-4}$, matched to the expected average wind optical depth throughout the bright phase of the outburst. Below, we calculate the self-consistent $\cal C$ as a function of time and see how the ourburst changes. Figure\,\ref{fig:206a} shows the outburst lightcurve with mass loss from the thermal wind taken into account. For comparison, the outburst lightcurve without the wind mass loss is shown as a grey line in the top panel of Fig.\,\ref{fig:206a}. Both lightcurves are virtually identical and we had to shift this grey line by 0.5 days to make it visible on the plot. The wind has negligible impact on the outburst lightcurve. The total wind mass loss rate peaks near the transition between hard and soft state ($l=0.02$), where it is comparable to the mass accretion rate onto the black hole (dashed line, first panel). In the hard state, the Compton radius is located within the accretion disc (second panel) but the critical luminosity is above the accretion luminosity $L=\epsilon \dot{M}_{\rm in} c^2$ (third panel), so the wind is mostly in the ``gravity-inhibited'' regime (regions C in Fig.\,25 of \citealt{1996ApJ...461..767W}). In the soft state, the Compton radius is beyond $R_{\rm out}$ and $L_{\rm crit}>L$, so the wind is actually the very weak outflow from a non-isothermal corona (region D).

\begin{figure}
\begin{center}
\includegraphics[width=\linewidth]{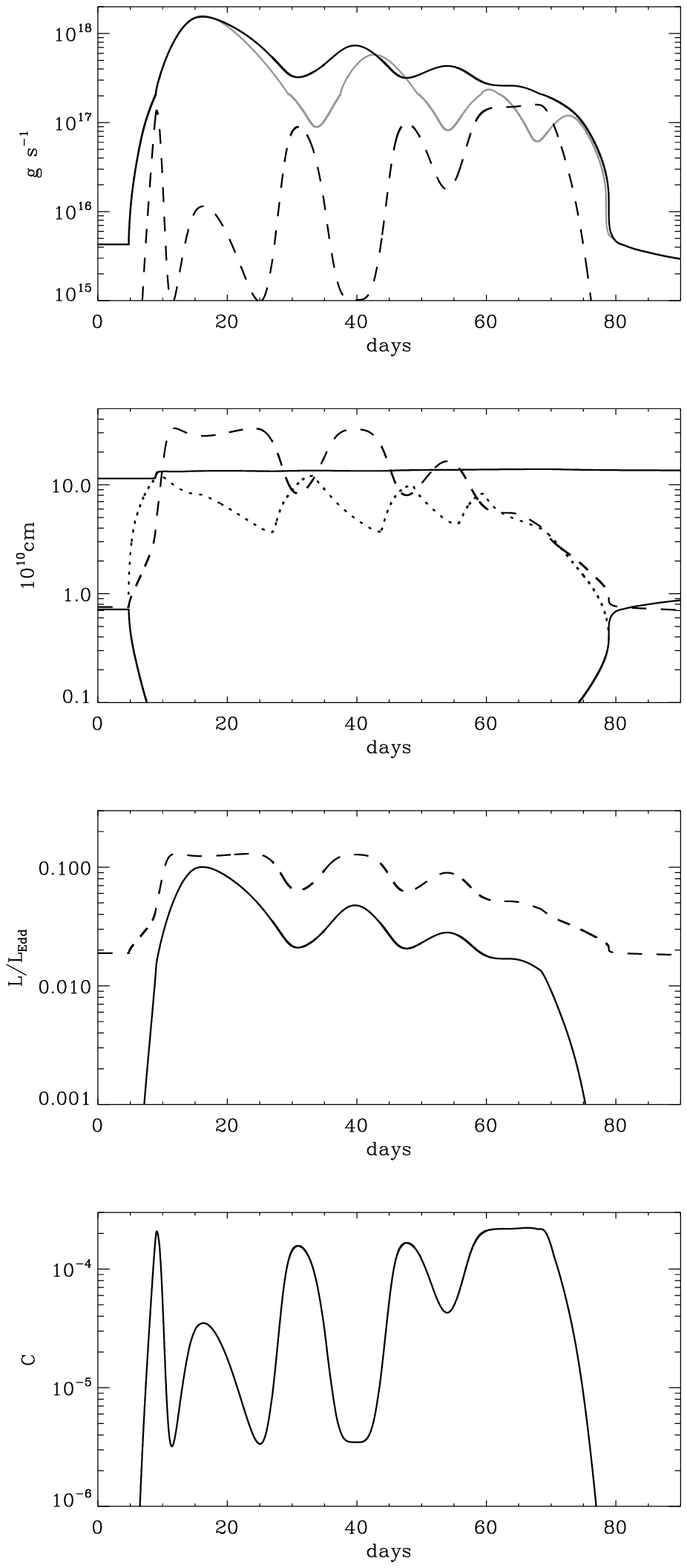} 
\caption{Same as Fig.\,\ref{fig:206a} except that $\cal C$ is calculated self-consistently from scattering in the wind (bottom panel, solid line). In the top panel, the grey solid line shows $\dot{M}_{\rm in}$ for a non-irradiated disc, with wind mass loss taken into account. All other line codes as in Fig.\,\ref{fig:206a}.}
\label{fig:206b}
\end{center}
\end{figure}

The irradiation parameter $\cal C$ estimated from scattering in the outflowing corona is within an order-of-magnitude of the constant $\cal C$ we used, but shows important fluctuations around $l=0.02$. These can significantly impact the lightcurves even though the wind mass loss rate is not important dynamically. Indeed, Fig.\,\ref{fig:206b} shows the result of taking $\cal C$ self-consistently from the scattered fraction. The outburst evolution is initially similar to the constant ${\cal C}=10^{-4}$ case and starts to differ around $t=20\rm\,d$. The drop in $\cal C$ to values $< 10^{-5}$ strongly decreases the irradiation flux and this leads to reflares. Reflares occur when the surface density behind the cooling front becomes higher than the maximum $\Sigma_{\rm max}$ allowed on the cold branch. Irradiation usually quenches reflares as it inhibits the propagation of the front and leads to smaller $\Sigma$ as more mass is accreted \citep{2001A&A...373..251D}. The reflares are clearly identified as a change in the propagation direction of the front as it switches from an inward-going cooling front to an outward-going heating front (dotted line, second panel from top). The four reflare episodes are also apparent as modulations of the lightcurve. This is typical of unirradiated discs. To illustrate this, the grey line in the top panel of Fig.\,\ref{fig:206b} shows the lightcurve for an unirradiated disc (including mass loss from the wind). The lightcurve shows a similar sequence of reflares. This type of lightcurve is not observed so scattering in the wind, as modelled here, is insufficient to provide the required level of sustained irradiation heating during outburst in e.g. XTE J1118$+$480.

The outbursts are longer than for the models in Fig.\,\ref{fig:206a}. This is not a numerical issue. The outburst cycles that we show are always relaxed, meaning they are independent of the initial conditions, and with 3600 grid points they are resolution independent. The recurrence timescale does change between models but remains around 4700 days. One reason for the longer outburst is that the amount of mass accreted during outburst is small compared to the disc mass (the disc mass varies by $\approx 20$\% in all these models). Another reason is the presence of reflares. The back and forth movement of the front in a disc that remains massive certainly acts to lengthen the outburst.

\subsection{A long orbital period system}
We also consider an accretion disc around a $M=9\rm\,M_\odot$ black hole with a mass transfer rate $\dot{M}_{\rm t}=6\times 10^{16}\rm\,g\,s^{-1}$. The disc has a size $\approx 10^{12}\rm\,cm$, corresponding to an orbital period $\approx 6\rm\,d$, viscosity parameters  $\alpha_{\rm h}=0.2$ and $\alpha_{\rm c}=0.02$, and an irradiation parameter ${\cal C}=5\times 10^{-3}$. The system parameters were chosen to be close to those of V404 Cyg. This is close to the largest disc that our code can handle with the vertical structures we have available at present. The values of $\alpha$ ensured the amplitude of the outburst reached $\approx 0.2 L_{\rm Edd}$, large enough to have a strong wind without entering the optically thick regime. $\cal C$ was chosen to be close to the time-averaged value of the wind $\cal C$.

\begin{figure}
\begin{center}
\includegraphics[width=\linewidth]{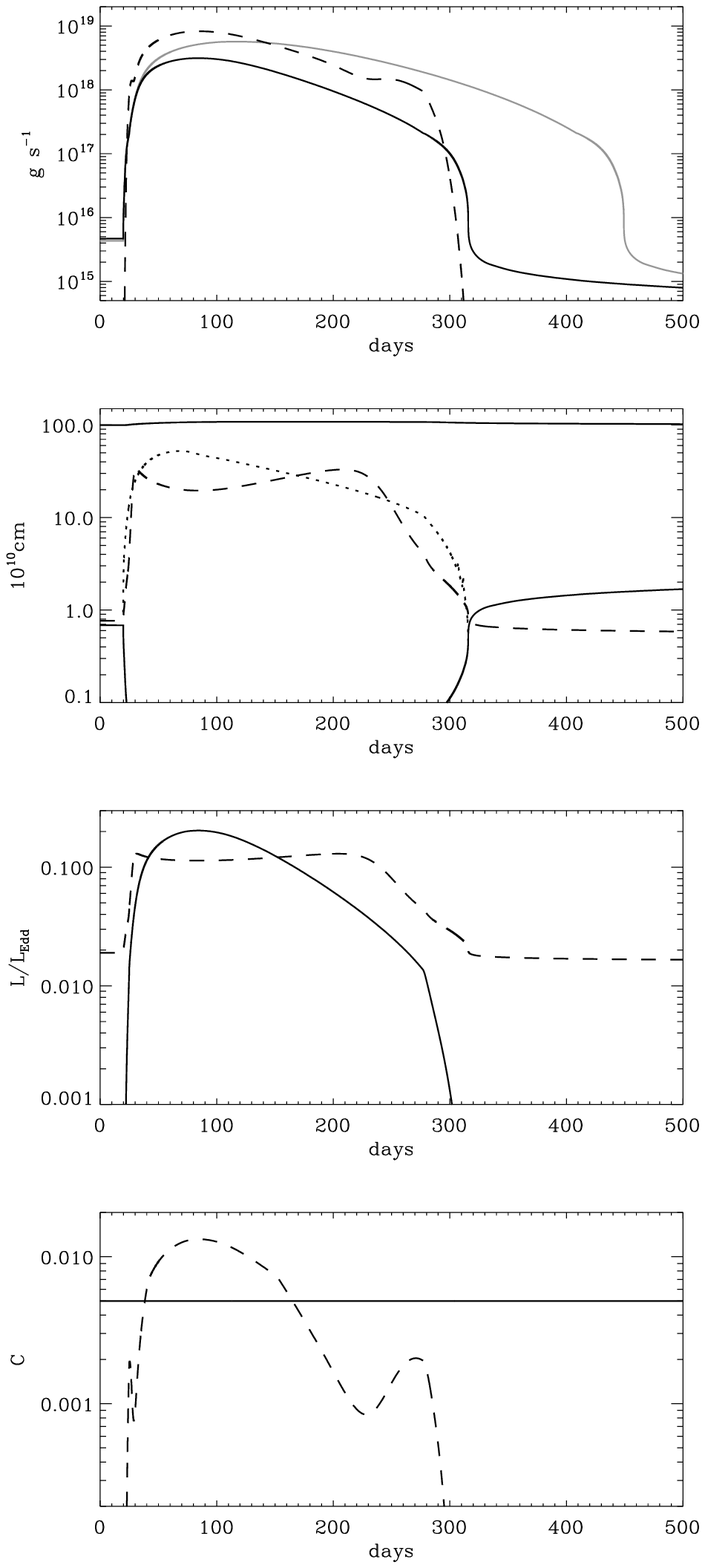} 
\caption{Outburst for a long orbital period black hole. Same description as Fig.\,\ref{fig:206a}, with the grey solid line in the top panel showing the case with no wind.}
\label{fig:404a}
\end{center}
\end{figure}

Figure\,\ref{fig:404a} shows the outburst lightcurve with mass loss from the thermal wind taken into account. For comparison, the outburst lightcurve without the wind mass loss is shown as a grey line in the top panel of Fig.\,\ref{fig:206a}. The mass accreted by the black hole is $\approx 40\%$ of the mass lost to the wind when integrated over the outburst duration: most of the mass is ejected from the system. The wind ends the outburst sooner for lack of material to keep the disc in the hot state.  It also decreases the peak luminosity  since $\dot{M}_{\rm in}$ is lowered by the mass lost in the outer disc (e.g. Fig.\,\ref{fig:stab}). Despite the shorter outburst more mass is accreted or ejected when the wind is turned on, so the recurrence time between outbursts accordingly increases from 75 (no wind) to 85 years (wind). 

Unlike the short orbital case described above, a bona fide Compton wind develops here because of the large disc. The Compton radius $0.2 R_{\rm IC}$ is always within the accretion disc during the outburst (second panel of Fig.\,\ref{fig:206a}) and the luminosity is higher than $L_{\rm crit}$ for at least part of the outburst (third panel). Thus, the nature of the outflow evolves during outburst between a ``steadily-heated free wind'' (region B in Fig.\,5 of \citealt{1996ApJ...461..767W}) and the Compton ``isothermal wind'' (region A) first described by \citet{1983ApJ...271...70B}. The switch between the hard and soft state is smoother than with the smaller accretion disc described in \S\ref{sporb}: the wind outflow rate is nearly proportional to $\dot{M}_{\rm in }$ throughout the outburst.  The decay is close to exponential, even though the front does not reach $R_{\rm out}$, with a decay timescale decreased by about 30\% compared to the case without wind. 

\begin{figure}
\begin{center}
\includegraphics[width=\linewidth]{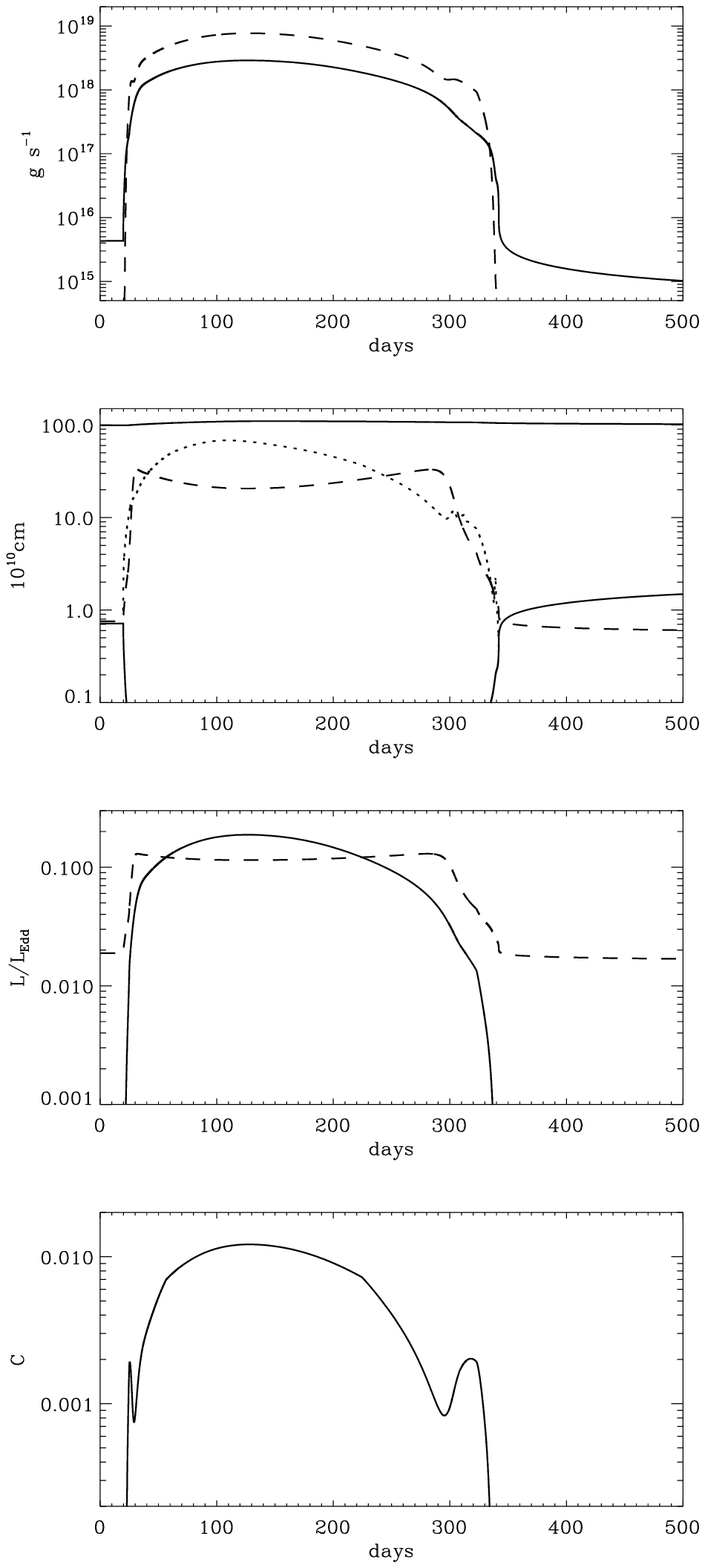} 
\caption{Same as  Fig.\,\ref{fig:404a} except for $\cal C$, which is calculated self-consistently from the scattered fraction in the wind.}
\label{fig:404b}
\end{center}
\end{figure}

The bottom panel of Fig.\,\ref{fig:404a} shows the estimated ${\cal C}$ from scattering. Its amplitude reaches $10^{-2}$, a factor 2 higher than the constant value assumed for Fig.\,\ref{fig:404a}. The two peaks in $\cal C$ at beginning and end of the outburst are associated with the hard to soft transition at $l=0.02$. The impact of the transition on $\dot{M}_{\rm w}$ is weaker than in \S\ref{sporb}, as noted above. Fig.\,\ref{fig:404b} shows how the outburst is modified by taking the self-consistent value of ${\cal C}$. The higher value reached by $\cal C$ lengthens slightly the outburst as irradiation maintains the disc in the hot state for a longer time. The evolution of $\cal C$ during the outburst moves the lightcurve to a more flat-top shape than a fast-rise exponential decay, reminiscent e.g. of the 1996 outburst of GRO J1655$-$40  \citep{2000A&A...354..987E}. GRO J1655$-$40 has a $7\rm\,M_\odot$ black hole with an average accretion rate of $\dot{M}_{\rm in}\approx 5\times 10^{16}\rm\,g\,s^{-1}$ and a disc size of $5\times 10^{11}\rm\,cm$ \citep{2012MNRAS.424.1991C}, reasonably close to the parameters of the model shown here.


\section{Discussion}

\subsection{The impact of thermal wind mass loss on lightcurves}
\begin{figure}
\begin{center}
\includegraphics[width=\linewidth]{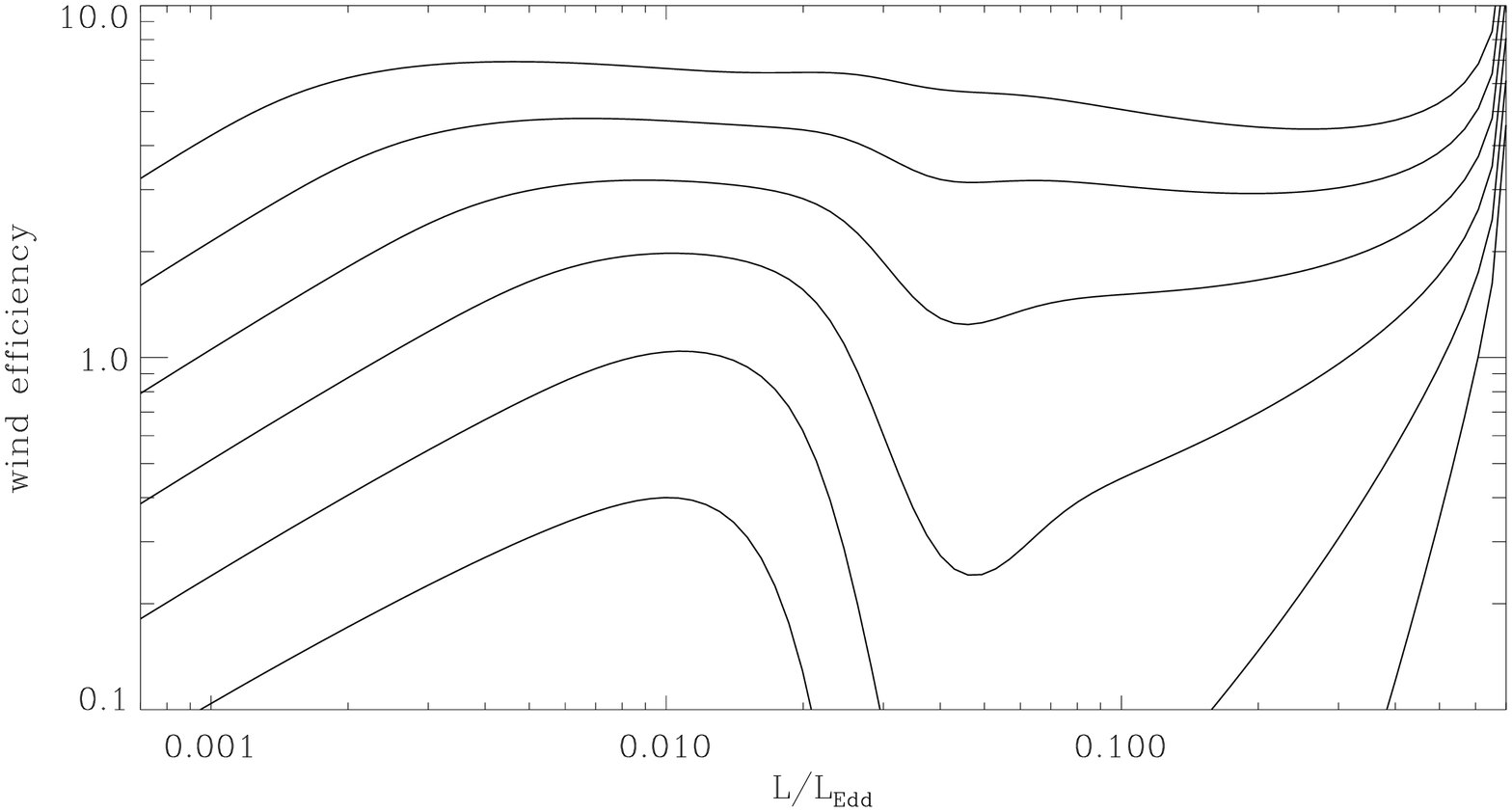} 
\caption{Compton wind efficiency $\eta_{\rm w}=\dot{M}_{\rm w}/\dot{M}$. The mass accretion rate is related to luminosity by $L=0.08 \dot{M}c^2$. The compact object is a 9$\rm\,M_\odot$ black hole. The outer disc radius goes as (from top to bottom) 32, 16, 8, 4, 2, 1 $\times 10^{11}\rm\,cm$, representative of the variations in X-ray binary disc sizes from short to long $P_{\rm orb}$.}
\label{fig:eff}
\end{center}
\end{figure}

The thermal wind shortens the outburst, as expected, but probably not enough to explain the rapid decay timescale of BHXB outbursts measured by  \citet{2018Natur.554...69T}. In the most extreme case that we have explored, shown in Fig.~\ref{fig:404a}, the outburst duration is shortened from 450 days to 300 days in our long orbital period system. The decay timescale that would be inferred from the lightcurve would lead to an effective $\alpha$ about 1.5 times greater than the value deduced from the lightcurve without a thermal wind. Achieving shorter outburst duration would require much more extreme mass loss rates much greater than we have in our model outburst. We show in Fig.\,\ref{fig:eff} the wind efficiency $\eta_{\rm w}$ as a function of the irradiation luminosity $L$ for our thermal wind model. The wind efficiency is defined as the ratio $\eta_{\rm w}=\dot{M}_{\rm w}/\dot{M}_{\rm in}$ with $L=0.08 \dot{M}_{\rm in} c^2$. The different curves correspond to different disc sizes. Fig.\,\ref{fig:eff} shows that $\eta_{\rm w}$ is unlikely to become much greater than $\approx 2$ to 5, except possibly close to the Eddington luminosity as we detail below. The model in Fig.~\ref{fig:404a} has $\eta_{\rm w}\approx 3$. In most systems, the effect on the lightcurve will be modest or negligible.

An extreme example of a rapid decay is V404 Cyg. The parameters of the long $P_{\rm orb}$ system were chosen to be close to those of V404 Cyg, whose last outburst lasted only a couple of weeks and showed a pronounced disc outflow, conjectured to be a thermal wind \citep{2016Natur.534...75M}.  The mass in the disc at the onset of the outburst is $\sim 3\times 10^{-7}\rm\,M_\odot$ in our models\footnote{This is much less than the maximum allowed mass $\sim 10^{-5}\rm M_\odot$ (Eq.\,54 of \citealt{Lasota:2001th} with $\alpha=0.02$) because $\Sigma$ is much lower than $\Sigma_{\rm max}$ in the outer disc, where most of the mass is stored.}. Blowing away most of this mass would require a sustained outflow rate of $\approx 10^{-5}\rm\,M_\odot\,yr^{-1}\approx  30\,\dot{M}_{\rm Edd}$ over 15 days for a 9\,M$_\odot$ black hole. Such very high mass outflow rates may be reached close to the Eddington luminosity as electron scattering contributes to the driving force of the wind. Figure\,\ref{fig:eff} shows the mass outflow rate in the wind diverges near $L\approx 0.7 L_{\rm Edd}$ due to the estimated radiation driving correction (Eq.\,\ref{eq:ric}). In principle, it might thus be possible to shorten the outburst of V404 Cyg to a couple of weeks by fine-tuning the model parameters to sample this high luminosity region. In support, observations of V404 Cyg do indicate the source likely reached $L_{\rm Edd}$ \citep{2016Natur.529...54K} and was enshrouded by rapidly varying Compton-thick outflowing material  \citep{2017A&A...602A..40S} with an estimated $M_{\rm w}\approx 4\times 10^{-6}\rm\,M_\odot$ lost to the wind \citep{2019MNRAS.tmp.1750C}. The lower effective gravity due to the high radiation should enhance the wind \citep{2002ApJ...565..455P} but the mass loss must saturate at some level as the outflow becomes optically thick. \citet{2019MNRAS.484.4635H} do not find a significantly enhanced $\dot{M}_{\rm w}$ near $L_{\rm Edd}$ in their radiation-hydrodynamic simulations of thermal winds, but these neglect radiative driving by electron scattering. If winds are boosted near Eddington, a puzzle is why GRS 1915+105 has not been affected as much as V404 Cyg despite its luminosity also being close to Eddington and its disc size even greater. If the short duration of the V404 Cyg was due to a thermal wind, then this wind likely required very specific conditions. Instead, we speculate that the angular momentum transport was instead dominated by the jet. The system likely stayed in the (very) bright hard state during the outburst, where it has a strong jet which is almost certainly coupled to the accretion flow via the magnetic fields and could be responsible for angular momentum transport through the hot flow in this state (e.g. \citealt{2006A&A...447..813F}).

The parametrisation we used predicts a wind in the hard state. Figure~\ref{fig:eff} shows that the wind can be stronger in the brightest hard states than in the dimmest soft states due to the higher $T_{\rm IC}$, decreasing $R_{\rm IC}$. The effect on the wind efficiency is particularly pronounced for small discs. This is apparently in contradiction with observations since wind outflow signatures are mostly seen during the soft state (see Section 1). It is usually argued that the wind is highly ionised due to the high radiation temperature, erasing potential spectral line signatures. Another possibility is that the wind is quenched in the hard state. Irradiation does not produce a smooth heated atmosphere as it becomes optically thick along the disc, self shielding the outer radii from illumination until the disc flare means it rises above the shadow cast by the inner atmosphere. The height of the inner atmosphere is larger for the higher $T_{\rm IC}$ of the hard state, so casts a longer shadow which can completely shield the outer disc \citep{1983ApJ...271...70B}. The wind outflow might thus be completely suppressed in the hard state \citep{2019arXiv190511763T}. In any case, the impact on the outburst lightcurve of wind mass loss during the hard state remains negligible.

\begin{figure}
\begin{center}
\includegraphics[width=\linewidth]{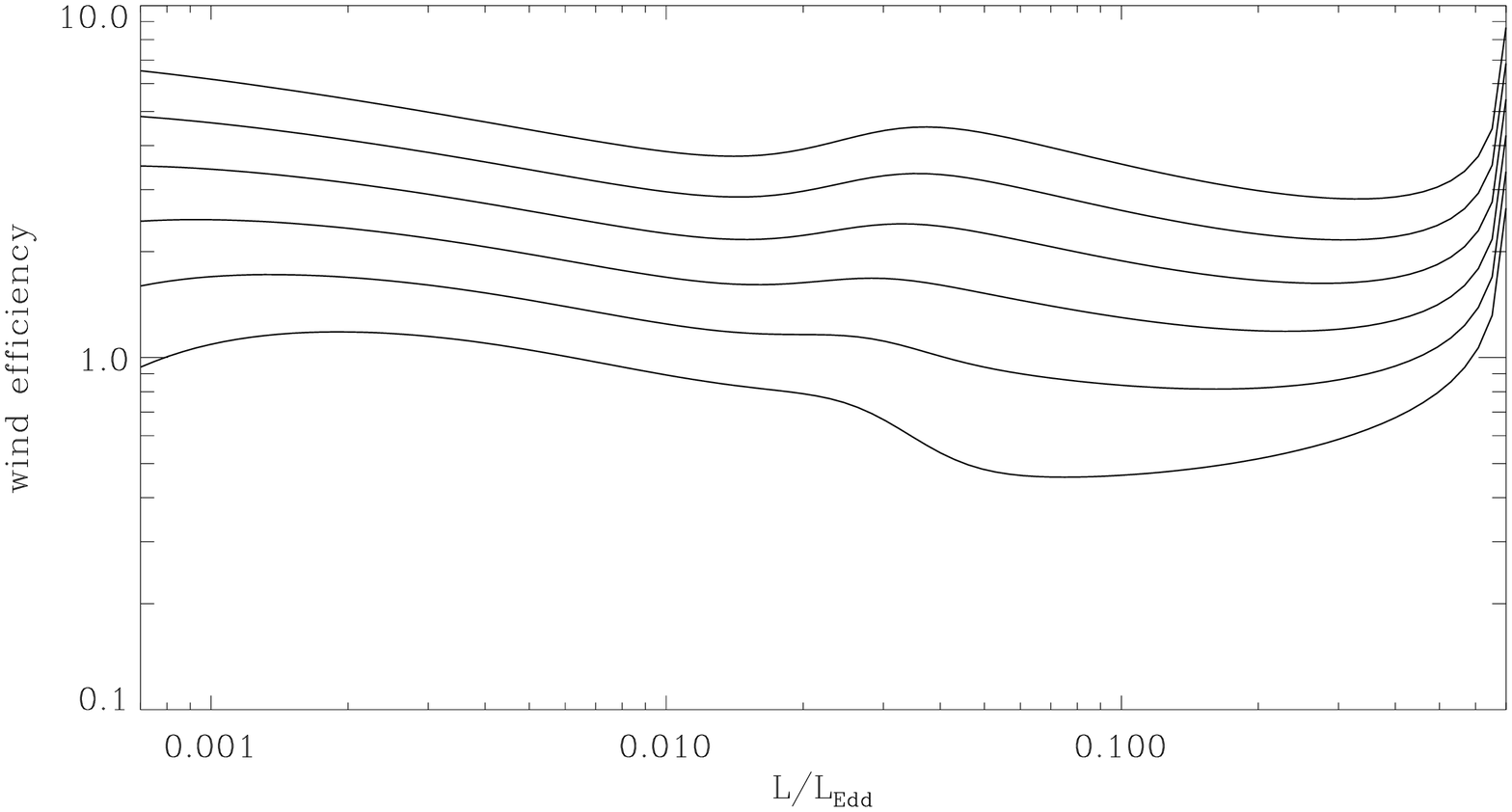} 
\caption{Same as Fig.\,\ref{fig:eff} with $T_{\rm IC}$ multiplied by a factor 10.}
\label{fig:eff10}
\end{center}
\end{figure}

The typical wind efficiency $\eta_{\rm w}\approx 2$ to 5 predicted for optically-thin Compton winds are also too low to enter the regime $\eta_{\rm w}\ga 15$ where an unstable cycle of inflowing/outflowing matter might exist \citep{1986ApJ...306...90S}. In practice, even the largest discs in BHXBs do not generate such strong outflows (Fig.\,\ref{fig:eff}). Uncertainties in the Compton temperature, caused e.g. by the very high state which is not included in our outburst model linking $L/L_{\rm Edd}$ to $T_{\rm IC}$ but which has both strong disc emission and strong but steep power law tail, are unlikely to change this. Figure~\ref{fig:eff10} shows that arbitrarily multiplying $T_{\rm IC}$ by a factor 10 in Eq.\,\ref{eq:tic} changes the variations with $L$ (because $R_{\rm IC}$ is smaller) without changing the wind efficiency\footnote{Figure~\ref{fig:eff10} also shows the drop in wind efficiency at the state transition is much smaller when a higher $T_{\rm IC}$ is assumed. This would lead to a smaller jump in the stability line (see Section \ref{sec:stab}).}. This also indicates that changing the luminosity for the switch from the hard to the soft state will not change the maximum value of $\eta_{\rm w}$ much.  For the same reason, taking into account the hysteresis pattern in the hardness -- intensity diagram is unlikely to impact our results. Hence, in as much as this parametrisation adequately reproduces the results of more complex simulations of the irradiation heating process \citep{1996ApJ...461..767W,2018MNRAS.473..838D,2019MNRAS.484.4635H,2019arXiv190511763T}, BHXBs lightcurves are unlikely to provide very strong dynamical signatures for the presence of mass loss due to thermal winds.

\subsection{The X-rays scattered in the wind}

The most important effect of thermal winds on the lightcurves is rather that they might change the fraction of X-rays scattered onto the disc. Compton winds are driven by X-ray heating of the upper layers of the outer disc, which must see the X-rays from the inner disc. The energy deposited by irradiation in the outer disc is also necessary to explain the optical luminosity of BHXB, their stability properties and their outburst lightcurves (Section 1). It has now long been known that disc flaring is insufficient to geometrically capture a sufficient amount of the (point-like) inner X-ray luminosity. In an unstable disc undergoing outbursts, the cold outer disc necessarily has a smaller scale height than the hot inner disc, which self-shields the outer disc from the X-ray irradiation \citep{dubus1999}. A wind, or a corona, such as one that can be provided by irradiation heating, can help scatter X-ray irradiation from the inner regions back to the outer disc. The launching of a thermal wind and the geometry of X-ray irradiation are thus directly connected. This scattered fraction can in principle be calculated self-consistently from simulations and may lead to interesting interplay, such as the suppression of the hard state thermal wind discussed above. It may supplement or dominate over direct irradiation of the flaring disc.

Following \citet{2019MNRAS.482..626K}, we find that a very simple  order-of-magnitude estimate of the scattered fraction $\cal C$ leads to values that can be dynamically significant, more so than the wind mass loss. Our stationary model (Section 3.1) shows strong outflows are required to change the disc temperature. This can be easily compensated by a relatively small change in $\cal C$ due to a slight change in wind density or geometry. Our short $P_{\rm orb}$ model demonstrates that our rough self-consistent evolution of $\cal C$ impacts the lightcurve much more than the wind outflow (Section 3.2). The outburst is very different than what is predicted when $\cal C$ is assumed to be constant. Although the scatter $\cal C$ can reach high values, the wind is not strong enough to maintain these high values and the lightcurve is essentially that of an unirradiated disc (Fig.\,\ref{fig:206b}). In the long $P_{\rm orb}$ system, the estimated $\cal C$ becomes somewhat higher than the typical values usually assumed and changes the lightcurve to a plateau-like state (Fig.\,\ref{fig:404b}). The order-of-magnitude estimate clearly breaks down for higher $L$ since the wind becomes opaque. 

Changes in $\cal C$ are directly related to variations in $\dot{M}_{\rm w} R_{\rm in}^{-1/2}$ if the wind speed is roughly proportional to the escape velocity at the inner edge of the thermal wind launching region (Eq.\,\ref{eq:estC}, see also Figs.\,\ref{fig:206b} and \ref{fig:404b}). Other mechanisms may also lead to changes in $\cal C$, such as warps \citep{2000ApJ...532.1069E,2000A&A...354..987E}. Tracking the time series evolution of $\cal C$ throughout an outburst using a combination of X-ray and optical data might thus identify times of important changes in wind properties, possibly relating them to changes in the spectrum, luminosity, column density, ionisation etc (Tetarenko et al., in prep.).

\section{Conclusion}
Compton winds are a natural outcome of X-ray irradiation of large accretion discs in BHXBs. The mass outflow rate in the wind is typically several times the accretion rate, which does shorten somewhat the outburst in long $P_{\rm orb}$ binaries. We do not find substantially new behaviour in the lightcurves due to the thermal wind mass loss, such as glitches or re-brightenings, at least under the assumptions used here to parametrise the wind. However, we do find that the lightcurves and stability properties are significantly affected when X-ray irradiation heating is derived from the estimated scattered light in the thermal wind. Scattering in the thermal wind couples the irradiation constant $\cal C$ to the disc size and the mass accretion rate. Stability can be achieved at lower mass accretion rates at long $P_{\rm orb}$ because $\cal C$ is then very high (Fig.\,\ref{fig:stab}). There are no remaining free parameters once $T_{\rm IC}$ is specified. The resulting stability curves separate persistent and transient systems very well in the $(P_{\rm orb},\dot{M}_{\rm in})$ plane, arguably better so than irradiated discs with $\cal C$ fixed at some identical value for all systems. 

We find that Compton wind mass loss is unlikely to be high enough to explain the rapid decay of a variety of BHXB outbursts in different spectral states \citep{2018Natur.554...69T}, though we cannot yet simulate the brightest outbursts. Instead, an alternative 
explanation of  those fast decays is angular momentum losses from a magnetic wind threading the disc and/or hot flow. This is a very effective way to change the outburst dynamics as shown for dwarf novae by \citet{2019A&A...626A.116S}, even when a negligible amount of matter is lost to the magnetic wind. Whereas thermal winds extract matter but no angular momentum, magnetic winds do exactly the opposite. \citet{2018MNRAS.481.2628W} find strong magnetic fields tend to inhibit thermal winds since the outflowing material, channeled by the field, is not able to be accelerated. One or the other may dominate at different times depending on where magnetic fields originate and diffuse through the disc in outburst. Understanding the latter is a difficult but an increasingly pressing issue.

\begin{acknowledgement}
We thank Micka\"el Coriat for providing the X-ray binary data used in Fig.\,\ref{fig:stab} and the participants of the ``Outflows 2019'' workshop held in Amsterdam for useful discussions. GD acknowledges support from {\it Centre National d'Etudes Spatiales} (CNES). BET acknowledges support from the University of Michigan through the McLaughlin Fellowship. CD acknowledges the Science and Technology Facilities Council (STFC) 
through grant ST/P000541/1 for support.
\end{acknowledgement}

\bibliographystyle{aa}
\bibliography{thermalwind}
\end{document}